# A note on dimensions of polynomial size circuits[*]

Xiaoyang Gu
Department of Computer Science
Iowa State University
Ames, IA 50011, USA
xiaoyang@cs.iastate.edu

**Abstract**

In this paper, we use resource-bounded dimension theory to investigate polynomial size circuits. We show that for every $i \geq 0$, P/poly has $i$th order scaled $p_3$-strong dimension 0. We also show that P/poly$^{\text{i.o.}}$ has $p_3$-dimension $1/2$, $p_3$-strong dimension 1. Our results improve previous measure results of Lutz (1992) and dimension results of Hitchcock and Vinodchandran (2004).

## 1 Introduction

Circuit-size complexity is one of the most investigated topics in computer science. In particular, much effort has been centered on the relationship between polynomial size circuits and uniform complexity classes. Kannan showed that ESPACE $\nsubseteq$ P/poly$^{\text{i.o.}}$ [9], i.e., that there exists a language in ESPACE that does not have polynomial size circuits even on only infinitely many lengths.

Lutz invented resource-bounded measure [11] as a powerful tool to examine the quantitative structure within complexity classes and quantified Kannan's separation result as

$$\mu(\text{P/poly}^{\text{i.o.}}|\text{ESPACE}) = 0,$$

which means that it is typical for a language in ESPACE not to have polynomial size circuits even on only infinitely many lengths. In the same paper, Lutz showed that for all $c > 0$,

$$\mu(\text{SIZE}^{\text{i.o.}}(n^c)|\text{EXP}) = \mu_{p_2}(\text{SIZE}^{\text{i.o.}}(n^c)) = 0$$

and

$$\mu(\text{P/poly}^{\text{i.o.}}|\text{E}_3) = \mu_{p_3}(\text{P/poly}^{\text{i.o.}}) = 0.$$

Measure theory is known not to distinguish among measure 0 sets. In classical analysis, Hausdorff dimension [4] and packing dimension [16, 15] serve as refined measurements that complement this limitation of measure. In computational complexity, Lutz et al. effectivized them as the resource-bounded dimension and strong dimension to examine the structure inside resource-bounded measure 0 sets [13, 3]. Very soon after the effectivization, the dimensions are further generalized to scaled dimensions to reveal the subtle relationship that cannot be addressed without scaling [7]. At the same time, resource-bounded dimension and strong dimension for individual sequences are defined to measure the level of randomness for individual sequences.

---

[*]This research was supported in part by a faculty startup grant of Pavan Aduri and National Science Foundation Grants 9988483 and 0344187.

Hitchcock and Vinodchandran [8] recently improved it with a dimension measurement of P/poly. They proved that for all $c > 0$,

$$\dim(\text{SIZE}(n^c)|\text{EXP}) = \dim_{\text{p}_2}(\text{SIZE}(n^c)) = 0$$

and

$$\dim(\text{P/poly}|\text{E}_3) = \dim_{\text{p}_3}(\text{P/poly}) = 0.$$

Recent result by Allender et al. [1, 2] regarding time-bounded Kolmogorov complexity KT and circuit size complexity of strings enables us to measure the class of polynomial size circuits more precisely. In section 4, we use the equivalence between KT complexity and circuit-size complexity to show that

$$\dim_{\text{p}_3}(\text{P/poly}^{\text{i.o.}}) = \dim(\text{P/poly}^{\text{i.o.}}|\text{E}_3) = 1/2$$

improving corresponding result by Lutz in [11]. Additionally, we show

$$\text{Dim}_{\text{p}_3}(\text{P/poly}^{\text{i.o.}}) = \text{Dim}(\text{P/poly}^{\text{i.o.}}|\text{E}_3) = 1.$$

We also improve a recent result by Hitchcock and Vinodchandran [8] from dimension to scaled strong dimension by showing that for all $i \in \mathbb{N}$,

$$\dim_{\text{p}_3}^{(i)}(\text{P/poly}) = \text{Dim}_{\text{p}_3}^{(i)}(\text{P/poly}) = 0.$$

Section 2 contains preliminaries. Section 3 is a review of some concepts and properties of resource-bounded measures and dimensions.

## 2 Preliminaries

In this paper, *languages* are sets of finite binary strings, i.e., subsets of $\{0, 1\}^*$. The empty string is denoted by $\lambda$. The length of a string $w$ is $|w|$. $|\lambda| = 0$. We fix a standard enumeration of all strings as $s_0 = \lambda$, $s_1 = 0$, $s_2 = 1$, $s_3 = 00$, etc. $\mathbf{C}$ is *Cantor space*, i.e., $\{0, 1\}^\infty$. $[\![\cdot]\!]$ is the boolean evaluation function. For a language $A$, we also identify it with its characteristic sequence $\chi_A \in \mathbf{C}$ such that $\chi_A = [\![s_0 \in A]\!][\![s_1 \in A]\!][\![s_2 \in A]\!]\cdots$. We use $A$ for $\chi_A$ in this paper. So $\mathbf{C}$ is the set of all languages. For integers $0 \leq i, j < |w|$, $w[i..j] = w[i]w[i+1]\cdots w[j]$ and $\lambda$ if $j < i$. We use the same convention to identify a finite consecutive part of a sequence also. If string $x$ is prefix of string $y$, we write $x \sqsubseteq y$. A string $w$ is a prefix of a sequence $S$, we write $w \sqsubseteq S$. $\otimes$ is the bitwise AND operator. In terms of languages, $\otimes$ corresponds to language intersection.

Regarding circuit size complexity, $\text{SIZE}(f(n)) = \{A \in \mathbf{C} | (\forall^\infty n) CS_A(n) \leq f(n)\}$, where $CS_A(n)$ is the number of gates in the smallest $n$-input Boolean circuit that decides $A \cap \{0, 1\}^n$. $\text{SIZE}^{\text{i.o.}}(f(n)) = \{A \in \mathbf{C} | (\exists^\infty n) CS_A(n) \leq f(n)\}$. For $x \in \{0, 1\}^*$, if $|x| = 2^k$ for some $k \in \mathbb{N}$, then define $\text{SIZE}(x)$ as the size of the smallest $k$ input circuit whose truth table is $x$. $\text{P/poly} = \bigcup_{c \in \mathbb{N}} \text{SIZE}(n^c)$ and $\text{P/poly}^{\text{i.o.}} = \bigcup_{c \in \mathbb{N}} \text{SIZE}^{\text{i.o.}}(n^c)$.

Let $s$ be a time constructible function. $\text{DTIME}(s)$ is the class of languages decidable in time $O(s)$ by deterministic Turing machines and $\text{DTIMEF}(s)$ is the class of functions computable in time $O(s)$ by deterministic Turing transducers. $\text{DSPACE}(s)$ and $\text{DSPACEF}(s)$ are defined similarly. $\Delta$ represents a function class which serves as a resource bound. In this paper, $\Delta$ may be one of the following: pspace = DSPACEF($n^{O(1)}$), $\mathrm{p}_2$ = DTIMEF($2^{(\log n)^{O(1)}}$) = DTIMEF($n^{(\log n)^{O(1)}}$), and $\mathrm{p}_3$ = DTIMEF($2^{2^{(\log \log n)^{O(1)}}}$). Lutz defined resource bounded constructors [10, 11, 13] that generate complexity classes. For a resource bound $\Delta$ the corresponding class is denoted as $R(\Delta)$. The correspondance between resource bounds and complexity classes that we use in this paper includes $R(\mathrm{p}_2)=\mathrm{E}_2=\mathrm{EXP}$, $R(\mathrm{pspace})=\mathrm{ESPACE}$ and $R(\mathrm{p}_3)=\mathrm{E}_3=\mathrm{DTIME}(2^{2^{(\log n)^{O(1)}}})$.

## 3 Measures and Dimensions

In this section, we summarize some concepts and theorems about measures and dimensions that we will use in the development of our results.

**Definition.** For $s \in [0, \infty)$, an *s-supergale* is a function $d : \{0,1\}^* \to [0, \infty)$ such that for all $w \in \{0,1\}^*$

$$2^s d(w) \geq d(w0) + d(w1). \tag{3.1}$$

A *supermartingale* is a 1-supergale with equality in (3.1). The *success set* of a $s$-supergale $d$ is

$$S^\infty[d] = \{S \in \mathbf{C} | \limsup_{n \to \infty} d(S[0..n-1]) = \infty\}.$$

We say $d$ *succeeds* on $S \in \mathbf{C}$, if $S \in S^\infty[d]$. The *strong success set* of $d$ is

$$S^\infty_{\mathrm{str}}[d] = \{S \in \mathbf{C} | \liminf_{n \to \infty} d(S[0..n-1]) = \infty\}.$$

We say $d$ *succeeds strongly* on $S \in \mathbf{C}$, if $S \in S^\infty_{\mathrm{str}}[d]$.

Resource-bounded measures and dimensions are defined by enforcing resource bound on the computation of martingales.

**Definition.** A function $f : \{0,1\}^* \to \mathbb{R}$ is $\Delta$-*computable*, if there is a function $\hat{f} : \{0,1\}^* \times \mathbb{N} \to \mathbb{Q}$ such that $\hat{f} \in \Delta$ and for all $w \in \{0,1\}^*$ and $r \in \mathbb{N}$, $|\hat{f}(w, r) - d(w)| \leq 2^{-r}$.

**Definition.** Let $X \subseteq \mathbf{C}$. $X$ has $\Delta$-*measure* $0$ and we write $\mu_\Delta(X) = 0$, if there exists a $\Delta$-computable supermartingale $d$ such that $X \subseteq S^\infty[d]$. $X$ has $\Delta$-*measure* $1$, if $X^c$ has $\Delta$-measure $0$. $X$ has *measure* $0$ *in* $R(\Delta)$, if $\mu_\Delta(X \cap R(\Delta)) = 0$. $X$ has *measure* $1$ *in* $R(\Delta)$, if $\mu_\Delta(X^c \cap R(\Delta)) = 0$.

**Definition ([11, 12, 13, 14, 3]).** Let $X \subseteq \mathbf{C}$. The $\Delta$-*dimension* of $X$ is

$$\dim_\Delta(X) = \inf\{s \in [0, \infty) | X \subseteq S^\infty[d] \text{ for some } \Delta\text{-computable } s\text{-supergale } d\}.$$

The $\Delta$-*strong dimension* of $X$, denoted $\mathrm{Dim}_\Delta(X)$, is defined similarly with respect to strong success. The *dimension* of $X$ in $R(\Delta)$ is $\dim(X|R(\Delta)) = \dim_\Delta(X \cap R(\Delta))$. The *strong dimension* of $X$ in $R(\Delta)$ is $\mathrm{Dim}(X|R(\Delta)) = \mathrm{Dim}_\Delta(X \cap R(\Delta))$.

When $\Delta$ is the set of all functions (with no computational restriction), the above definition gives us the classical Hausdorff dimension $\dim_\mathrm{H}$ and packing dimension $\dim_\mathrm{P}$.

**Observation 3.1 (Lutz [11, 12, 13, 14], Athreya et al. [3]).** *1. For all $X \subseteq \mathbf{C}$ and all resource bounds $\Delta$, if $\dim_\Delta(X) < 1$ then $\mu_\Delta(X) = 0$.*

*2. For all $X \subseteq \mathbf{C}$ and all resource bounds $\Delta$, $\dim_\Delta(X) \leq \mathrm{Dim}_\Delta(X)$.*

3. For all $X \subseteq Y$ and all resource bounds $\Delta$, $\dim_\Delta(X) \leq \dim_\Delta(Y)$.

4. Let $\Delta, \Delta'$ be resource bounds such that $\Delta \subseteq \Delta'$. Then for all $X \subseteq \mathbf{C}$, $\dim_{\Delta'}(X) \leq \dim_\Delta(X)$.

In contrast to the classical measure and dimension theory, when resource bounds are enforced on computation of supergales, there are individual sequences that are not in the success set of any supergales. Therefore, randomness and dimension of individual sequences can now be defined.

**Definition.** Let $S \in \mathbf{C}$ be an infinite binary sequence. The $\Delta$-*dimension* of $S$, $\dim_\Delta(S) = \dim_\Delta(\{S\})$. The $\Delta$-*strong dimension* of $S$ is $\mathrm{Dim}_\Delta(S) = \mathrm{Dim}_\Delta(\{S\})$.

Hitchcock, Lutz, Mayordomo also introduced a theory of resource-bounded scaled dimension that has more distinguishing power [7, 6].

**Definition (Hitchcock, Lutz, Mayordomo [7]).** A *scale* is a continuous function $g : H \times [0, \infty) \to \mathbb{R}$ such that $H = (a, \infty)$ for some $a \in \mathbb{R} \cup \{-\infty\}$, $g(m, 1) = m$ for all $m \in H$, $g(m, 0) = g(m', 0) \geq 0$ for all $m, m' \in H$, for every sufficiently large $m \in H$, the function $s \mapsto g(m, s)$ is nonnegative and strictly increasing, and for all $s' > s \geq 0$, $\lim_{m \to \infty}[g(m, s') - g(m, s)] = \infty$.

**Definition (Hitchcock, Lutz, Mayordomo [7]).** Let $g : H \times [0, \infty) \to \mathbb{R}$ be a scale, and let $s \in [0, \infty)$. A *$g$-scaled $s$-supergale* ($s^{(g)}$-supergale) is a function $d : \{0, 1\}^* \to [0, \infty)$ such that for all $w \in \{0, 1\}^*$ with $|w| \in H$,

$$d(w) \geq 2^{-\Delta g(|w|, s)}[d(w0) + d(w1)], \tag{3.2}$$

where $\Delta g(m, s) = g(m+1, s) - g(m, s)$.

The definitions for scaled dimensions are identical to those of regular dimensions except the use of scaled supergales. In corresponding notations, we use superscript $^{(g)}$ to indicate the scale as in $\dim_\Delta^{(g)}(\cdot)$, $\mathrm{Dim}_\Delta^{(g)}(\cdot)$.

In this paper, we use the scales defined by Hitchcock, Lutz and Mayordomo in [7].

**Definition (Hitchcock, Lutz, Mayordomo[7]).** Let $g : H \times [0, \infty) \to \mathbb{R}$ be a scale.

1. The first rescaling of $g$ is the scale $g^\# : H^\# \times [0, \infty) \to \mathbb{R}$ defined by

$$H^\# = \{2^m | m \in H\}$$
$$g^\#(m, s) = 2^{g(\log m, s)}.$$

2. For each $i \in \mathbb{N}$, $a_0 = -\infty$, $a_{i+1} = 2^{a_i}$,

3. For each $i \in \mathbb{N}$, the $i$th scale $g_i : (a_{|i|}, \infty) \times [0, \infty) \to \mathbb{R}$ is defined such that

   (a) $g_0(m, s) = sm$.
   (b) For $i \geq 0$, $g_{i+1} = g_i^\#$.

When these scales are used, we use superscript $^{(i)}$ instead of $^{(g_i)}$. Resource-bounded 0th scaled dimensions and strong dimensions coincide with the regular dimensions and strong dimensions. With the scales defined above, it was shown that the scaled dimensions exhibit the following monotonicity with respect to the index of the scale.

**Theorem 3.2 (Hitchcock, Lutz, Mayordomo [7]).** *Let $i \in \mathbb{N}$ and $X \subseteq \mathbf{C}$. If $\dim_\Delta^{(i+1)}(X) < 1$, then $\dim_\Delta^{(i)}(X) = 0$.*

# 4 Polynomial Size Circuits

Lutz proved the following theorem regarding polynomial size circuits [11].

**Theorem 4.1.** *For all $c > 0$,*

$$\mu(\text{SIZE}^{\text{i.o.}}(n^c)|\text{EXP}) = \mu_{\text{p}_2}(\text{SIZE}^{\text{i.o.}}(n^c)) = 0$$

*and*

$$\mu(\text{P/poly}^{\text{i.o.}}|\text{E}_3) = \mu_{\text{p}_3}(\text{P/poly}^{\text{i.o.}}) = 0.$$

This result is recently improved by Hitchcock and Vinodchandran to dimension [8].

**Theorem 4.2.** *For all $c \geq 1$,*

$$\dim(\text{SIZE}(n^c)|\text{EXP}) = \dim_{\text{p}_2}(\text{SIZE}(n^c)) = 0$$

*and*

$$\dim(\text{P/poly}|\text{E}_3) = \dim_{\text{p}_3}(\text{P/poly}) = 0.$$

In this section, we use the relationship between time bounded Kolmogorov complexity and circuit complexity to give a more thorough discussion about the dimension of polynomial size circuits.

**Definition (Allender [1]).** Let $U$ be a universal Turing machine. Define $\text{KT}(x)$ to be

$$\min\{|p| + t| \text{ for all } i \leq |x|, U(p, i) = x_i \text{ in at most } t \text{ steps}\}.$$

**Theorem 4.3 (Allender et al. [1], [2]).** $\text{SIZE}^A(x) = O(\text{KT}^A(x)^3)$, *and* $\text{KT}^A(x) = O(\text{SIZE}^A(x) \cdot (\log \text{SIZE}^A(x) + \log |x|))$.

**Lemma 4.4.** *Let $A \subseteq \{0,1\}^*$.*

1. *$A \in \text{P/poly}^{\text{i.o.}}$ if and only if for some integer $c \in \mathbb{N}$, $\text{KT}(A[2^n - 1..2^{n+1} - 2]) \leq n^c$ for infinitely many $n \in \mathbb{N}$.*

2. *$A \in \text{P/poly}$ if and only if for some integer $c \in \mathbb{N}$, $\text{KT}(A[2^n - 1..2^{n+1} - 2]) \leq n^c$ for all but finitely many $n \in \mathbb{N}$.*

*Proof.* Both follow from Theorem 4.3. □

Using this Lemma, we first establish the following two theorems for individual language against $\text{P/poly}^{\text{i.o.}}$ and $\text{P/poly}$.

**Theorem 4.5.** *Let $A \subseteq \{0,1\}^*$ be a language such that $\dim_{\text{p}_2}(A) > 1/2$. Then $A \notin \text{P/poly}^{\text{i.o.}}$.*

*Proof.* Assume the hypothesis and suppose $A \in \text{P/poly}^{\text{i.o.}}$, then by Lemma 4.4, $\text{KT}(A[2^n - 1..2^{n+1} - 2]) < n^c$ for infinitely many $n$ and some fixed constant $c$.

Let $r > 1/2$ be a polynomial-time computable real number. It suffices to show that there exists a $\text{p}_2$-computable $r$-supergale $d$ that succeeds on $A$.

For $i \geq 1$, let
$$C_i = \{x \in \{0,1\}^{2^i} \mid \text{KT}(x) < i^c\}$$

$$C_i^w = \{x \in C_i \mid w[2^i - 1..|w| - 1] \sqsubseteq x\}$$

and let $d_i$ be such that

$$d_i(w) = \begin{cases} 2^{(r-1)|w|} & |w| < 2^i \\ d_i(w[0..2^i - 2])2^{r(|w|-(2^i-1))}\frac{|C_i^w|}{|C_i|} & 2^i \leq |w| \leq 2^{i+1} - 1 \\ 2^{(r-1)(|w|-(2^{i+1}-1))}d_i(w[0..2^{i+1} - 2]) & |w| > 2^{i+1} - 1. \end{cases}$$

$d_i$ simulates the universal Turing machine $U$ to enumerate $C_i$ by cycling all programs of length up to $i^c$ and all bit indices less than or equal to $2^k$ within running time less than $i^c$. For every such program, a valid simulation generates $2^i$ bits and by concatenating them, we get a output string of length $2^i$ in $C_i$.

During the enumeration, $d_i$ counts the number of strings in $C_i$ and in $C_i^w$ to get $|C_i|$ and $|C_i^w|$. Note that $|C_i| \leq 2^{i^c}$.

Let $d = \sum_{i=1}^{\infty} \frac{1}{i^2} d_i$. It is easy to verify that $d$ is a $p_2$-computable $r$-supergale.

For $n \geq 1$ such that $\mathrm{KT}(A[2^n - 1..2^{n+1} - 2]) < n^c$,

$$\begin{aligned} d(A[0..2^{n+1} - 2]) &\geq \frac{1}{n^2} d_n(A[0..2^{n+1} - 2]) \\ &\geq \frac{1}{n^2} 2^{(r-1)(2^n-1)} 2^{r2^n} \frac{\left|C_n^{A[0..2^{n+1}-2]}\right|}{|C_i|} \\ &= \frac{2^{(2r-1)2^n - r + 1}}{n^2 2^{n^c}}. \end{aligned}$$

It is clear that when $n$ is sufficiently large, $d(A[0..2^{n+1} - 2])$ is unbounded. Since $\mathrm{KT}(A[2^n - 1..2^{n+1} - 2]) < n^c$ for infinitely many $n$, the value that the $r$-supergale $d$ can obtain on along $A$ is unbounded and thus $\dim_{p_2}(A) \leq r$. Since polynomial-time computable real numbers are dense, $\dim_{p_2}(A) \leq 1/2$, which contradicts with our assumption. Therefore, $A \notin \mathrm{P/poly}^{\mathrm{i.o.}}$. □

By using the above construction and changing the simulation from only for length $i$ strings to for strings of length up to $i$, we can establish an analogous result regarding P/poly, however, with strong dimension.

**Theorem 4.6.** *Let $A \subseteq \{0, 1\}^*$ be a language such that $\mathrm{Dim}_{p_2}(A) > 0$. Then $A \notin \mathrm{P/poly}$.*

*Proof.* Assume the hypothesis and suppose $A \in \mathrm{P/poly}$, then by Lemma 4.4, $\mathrm{KT}(A[2^n - 1..2^{n+1} - 2]) < n^c$ for all but finitely many $n \in \mathbb{N}$ for some fixed constant $c$.

Let $r > 0$ be a polynomial-time computable real number.

For $i \geq 1$, let

$$C_{\leq i} = \{x \in \{0,1\}^{2^{i+1}-1} \mid \mathrm{KT}(x[2^k - 1..2^{k+1} - 2]) < k^c, 0 < k \leq i\}$$

$$C_{\leq i}^w = \{x \in C_{\leq i} \mid w \sqsubseteq x\}$$

and let $d_i$ be such that

$$d_i(w) = \begin{cases} 2^{r|w|} \frac{|C_{\leq i}^w|}{|C_{\leq i}|} & |w| \leq 2^{i+1} - 1 \\ 2^{(r-1)(|w|-(2^{i+1}-1))} d_i(w[0..2^{i+1} - 2]) & |w| > 2^{i+1} - 1. \end{cases}$$

$d_i$ simulates the universal Turing machine $U$ to enumerate $C_{\leq i}$ by cycling all programs of length up to $k^c$ and all bit indices less than or equal to $2^k$ within running time less than $k^c$ for $k = 0, 1, \ldots, i$ in a depth first fashion. For every such $k$ and a particular program, a valid simulation generates $2^k$ bits and by concatenating them, we get a output string of length $2^k$. By concatenating the output for $k$ from 0 to $i$, we get a string of length $2^{i+1} - 1$ in $C_{\leq i}$.

During the enumeration, $d_i$ counts the number of strings in $C_{\leq i}$ and in $C_{\leq i}^w$ to get $|C_{\leq i}|$ and $|C_{\leq i}^w|$. Note that $|C_{\leq i}| \leq 2^{i^c+1}$.

Let $d = \sum_{i=1}^{\infty} \frac{1}{i^2} d_i$. It is easy to verify that $d$ is a $p_2$-computable $r$-supergale.

For $n > 1$ and $0 < k \leq 2^n$, we have

$$d(A[0..2^n - 2 + k]) \geq \frac{1}{n^2} d_n(A[0..2^n - 2 + k])$$
$$= \frac{1}{n^2} 2^{r(2^n - 1 + k)} \frac{\left|C_{\leq n}^{A[0..2^n-2+k]}\right|}{|C_{\leq n}|}$$
$$\geq \frac{1}{n^2} 2^{r(2^n - 1 + k)} \frac{1}{2^{n^c+1}}.$$

Therefore, when $n$ is large enough, the value of the $r$-supergale along $A$ goes to infinity regardless of the value of $k$, i.e.,

$$\liminf_{n \to \infty} d(A[0..n-1]) = \infty.$$

So the $r$-supergale $d$ succeeds strongly on $A$, and hence the $\text{Dim}_{p_2}(A) \leq r$. By the density of polynomial-time computable real numbers, $\text{Dim}_{p_2}(A) = 0$ contradicting our assumption. Hence, $A \notin \text{P/poly}$. $\square$

Here we give an observation on scaled dimension that simplifies the calculation of scaled dimensions.

**Observation 4.7.** *Let $g_1$, $g_2$ be two scales and $s_1, s_2 \in [0, \infty)$. Let $d : \{0,1\}^* \to [0, \infty)$ be a $g_1$-scaled $s_1$-supergale ($s_1^{(g_1)}$-supergale), i.e.,*

$$d(w) \geq 2^{\Delta g_1(|w|,s_1)}[d(w0) + d(w1)].$$

*Then function $d' : \{0,1\}^* \to [0, \infty)$ defined by $d'(w) = d(w) 2^{g_2(|w|,s_2) - g_1(|w|,s_1)}$ is an $s_2^{(g_2)}$-supergale.*

*Proof.* This observation follows from easy verification of the $s_2^{(g_2)}$-supergale condition (3.2). $\square$

This observation tells us that we may obtain scaled dimensions by developing a supergale and examining the rate at which the supergale succeeds on a sequence or a set of sequences. Lemma 4.2 in [6] is a similar result using the log-loss unpredictability characterization of dimension [5].

Now we can use Observation 4.7 to analyze the proof of Theorem 4.6 and obtain the following theorem regarding the scaled dimension of individual languages in P/poly.

**Theorem 4.8.** *Let $j \geq \mathbb{N}$ and $A \subseteq \{0,1\}^*$ be a language such that $\text{Dim}_{p_2}^{(j)}(A) > 0$. Then $A \notin \text{P/poly}$.*

*Proof.* Assume the hypothesis and suppose $A \in \text{P/poly}$, then by Lemma 4.4, $\text{KT}(A[2^n - 1..2^{n+1} - 2]) < n^c$ for all but finitely many $n \in \mathbb{N}$ for some fixed constant $c$. We use the same $d_i$ (for all $i \in \mathbb{N}$) defined in the proof of Theorem 4.6 and let $d = \sum_{i=1}^{\infty} \frac{1}{i^i} d_i$. It is clear that $d$ is a $\text{p}_2$-computable $r$-supergale.

Define $d'$ such that
$$d'(w) = d(w)2^{g_j(|w|,s) - g_0(|w|,r)}.$$

By Observation 4.7, $d'$ is a $\text{p}_2$-computable $s^{(j)}$-supergale and
$$\begin{aligned} d'(A[0..2^n - 2 + k]) &\geq \frac{1}{n^2} d_n(A[0..2^n - 2 + k]) \frac{2^{g_j(2^n - 1 + k, s)}}{2^{r(2^n - 1 + k)}} \\ &\geq \frac{1}{n^2} \frac{2^{r(2^n - 1 + k)}}{2^{n^c + 1}} \frac{2^{g_j(2^n - 1 + k, s)}}{2^{r(2^n - 1 + k)}} \\ &= \frac{2^{g_j(2^n - 1 + k, s)}}{n^2 2^{n^c + 1}} \end{aligned}$$

It is easy to verify that for all $s > 0$, $c \in \mathbb{N}$ and $0 < k \leq 2^n$, when $n$ is sufficiently large, $g_j(2^n - 1 + k, s)$ dominates $n^{c+2}$. So $\liminf_{n \to \infty} d'(A[0..2^n - 2 + k]) = \infty$, for all $s > 0$ and thus $\text{Dim}_{\text{p}_2}^{(j)}(A) = 0$, which contradicts with our assumption. □

By using standard techniques, we can obtain the following two corollaries improving Theorem 4.1.

**Corollary 4.9.** $\dim_{\text{p}_3}(\text{P/poly}^{\text{i.o.}}) \leq 1/2$.

**Corollary 4.10.** *For all* $i \in \mathbb{N}$, $\dim_{\text{p}_3}(\text{P/poly}) = \text{Dim}_{\text{p}_3}(\text{P/poly}) = \text{Dim}_{\text{p}_3}^{(i)}(\text{P/poly}) = 0$.

The proof of the above theorems also imply the following.

**Corollary 4.11.** *For all* $c > 0$
$$\dim_{\text{p}_2}(\text{SIZE}^{\text{i.o.}}(n^c)) \leq 1/2$$
*and for all* $i \in \mathbb{N}$
$$\dim_{\text{p}_2}(\text{SIZE}(n^c)) = \text{Dim}_{\text{p}_2}(\text{SIZE}(n^c)) = \text{Dim}_{\text{p}_2}^{(i)}(\text{SIZE}(n^c)) = 0.$$

Jack Lutz suggested that the upper bound for dimension in Corollary 4.9 is tight. Now we prove a general theorem on dimension lower bound of infinitely often classes, which implies this observation.

In the proof, we use the following Measure Conservation Theorem.

**Theorem 4.12 (Lutz [11]).** $R(\Delta)$ *does not have measure* $0$ *in* $R(\Delta)$.

**Theorem 4.13.** *Let* $\mathcal{C}$ *be a language class that is closed under intersection and contains the trivial language* $A = \emptyset$. *Then for all* $\Delta$, $\dim(\mathcal{C}^{\text{i.o.}}|R(\Delta)) \geq 1/2$ *and* $\text{Dim}(\mathcal{C}^{\text{i.o.}}|R(\Delta)) = 1$.

*Proof.* Let $L = \{w \in \{0,1\}^* \mid |w| \text{ is not a power of } 2\}$.

Let $X = \{B \otimes L | B \in R(\Delta)\}$. It is clear that $X \subseteq A^{\text{i.o.}} \subseteq \mathcal{C}^{\text{i.o.}}$.

Suppse $\dim(X|R(\Delta)) < 1/2$, i.e., for some $s < 1/2$, there exists a $\Delta$-computable $s$-supergale $d$ such that $X \subseteq S^{\infty}[d]$, i.e., for every $S \in X$,
$$\limsup_{n \to \infty} d(S[0..n - 1]) = \infty.$$

Let $d^{\otimes}(\cdot) \equiv d(\cdot \otimes L)$. For $w \in \{0,1\}^*$ and $b \in \{0,1\}$, define $d'$ using the following recursion.

$$\begin{cases} d'(\lambda) = d(\lambda) \\ d'(wb) = d'(w) & |s_{|w|}| \text{ is a power of 2} \\ d'(wb) = 2^{1-s} d'(w) \frac{d^{\otimes}(wb)}{d^{\otimes}(w)} & \text{otherwise.} \end{cases} \quad (4.1)$$

It is easy to verify that $d'$ is a $\Delta$-computable supermartingale.

Let $I = \{n \mid (\exists k \in \mathbb{N}) |s_n| = 2^k\}$.

Let $S \subseteq L$.

For $n - 1 \in I$,
$$d'(S[0..n-1]) = d'(S[0..n-2]).$$

Since $d$ is a $s$-supergale,
$$d'(S[0..n-1]) \leq 2^s d'(S[0..n-2]).$$

Thus
$$\frac{d'(S[0..n-1])}{d'(S[0..n-2])} = 1 \geq \frac{d'(S[0..n-1])}{2^s d'(S[0..n-2])}, \forall (n-1) \in I. \quad (4.2)$$

For $n - 1 \notin I$,
$$d'(S[0..n-1]) = 2^{1-s} d'(S[0..n-2]) \frac{d(S[0..n-1])}{d(S[0..n-2])}.$$

Thus
$$\frac{d'(S[0..n-1])}{d'(S[0..n-2])} = \frac{2d(S[0..n-1])}{2^s d(S[0..n-2])}, \forall (n-1) \notin I. \quad (4.3)$$

Let $\#n = |\{k \in \mathbb{N} \mid 0 < k \leq n-1 \text{ and } k \in I\}|$. It is easy to verify that
$$\#n \leq n/2 + 2\sqrt{n/2}.$$

By (4.2) and (4.3),

$$\begin{aligned}
d'(S[0..n-1]) &= d'(S[0..0]) \prod_{i=1}^{n-1} \frac{d'(S[0..i])}{d'(S[0..i-1])} \\
&\geq d(\lambda) \prod_{i=1}^{n-1} \left[ \frac{2d(S[0..i])}{2^s d(S[0..i-1])} [\![ i \notin I ]\!] + \frac{d(S[0..i])}{2^s d(S[0..i-1])} [\![ i \in I ]\!] \right] \\
&= d(S[0..n-1]) \left( \frac{2}{2^s} \right)^{n-1-\#n} \left( \frac{1}{2^s} \right)^{\#n} \\
&\geq d(S[0..n-1]) \frac{2^{n-1-n/2-2\sqrt{n/2}}}{2^{s(n-1)}}.
\end{aligned}$$

Note that for all $S \in \mathbf{C}$, $S \otimes L \subseteq L$. Thus for all sufficiently large $n$,
$$d'((S \otimes L)[0..n-1]) \geq d((S \otimes L)[0..n-1]).$$

Then for all $S \in \mathbf{C}$,
$$\limsup_{n \to \infty} d'((S \otimes L)[0..n-1]) \geq \limsup_{n \to \infty} d((S \otimes L)[0..n-1]).$$

For every $S \in X$, $S \otimes L = S$ and $X \subseteq S^\infty[d]$, so $X \subseteq S^\infty[d']$.

For $w, w' \in \{0,1\}^*$, by the construction of $d'$,

$$d'(w) \neq d'(w') \Rightarrow (\exists i \notin I) w[i] \neq w'[i]. \tag{4.4}$$

Note that for any $S \in \mathbf{C}$,

$$S[i] = (S \otimes L)[i], \forall i \notin I. \tag{4.5}$$

Let $B \in R(\Delta)$. $B \otimes L \in X$. So $B \otimes L \in S^\infty[d']$. By (4.4) and (4.5), $B \in S^\infty[d']$. Therefore $R(\Delta) \subseteq S^\infty[d']$, i.e., $\mu_\Delta(R(\Delta)) = 0$, which contradicts with the Measure Conservation Theorem (Theorem 4.12). Therefore, $\dim(X|R(\Delta)) \geq 1/2$ and by Observation 3.1, $\dim(\mathcal{C}^{\text{i.o.}}|R(\Delta)) \geq 1/2$.

Now we suppose $\text{Dim}(X|R(\Delta)) < 1$, i.e. for some $s < 1$, there exists a $\Delta$-computable $s$-supergale $d$ such that $X \subseteq S^\infty_{\text{str}}[d]$, i.e., for every $S \in X$,

$$\liminf_{n \to \infty} d(S[0..n-1]) = \infty.$$

Define $d'$ as in (4.1).

Now for every $n \in \mathbb{N}$ such that $n = 2 \cdot 2^{2^k-1} - 1$ for some $k \geq 1$,

$$\#n \leq \sqrt{n}.$$

So for sufficiently large $n$ such that $n = 2 \cdot 2^{2^k-1} - 1$ for some $k \in \mathbb{N}$ and every $S \subseteq L$,

$$d'(S[0..n-1]) \geq d(S[0..n-1]) \frac{2^{n-1-\#n}}{2^{s(n-1)}} \geq d(S[0..n-1]).$$

For $B \in R(\Delta)$, apply the argument for the $\Delta$ dimension case to those $n \in \mathbb{N}$ such that $n = 2 \cdot 2^{2^k-1} - 1$ for some $k \geq 1$ and we have $B \otimes L \in S^\infty[d']$, and $B \in S^\infty[d']$ and thus $R(\Delta) \subseteq S^\infty[d']$, which contradicts with the Measure Conservation Theorem. Thus, again by Observation 3.1, $\text{Dim}(\mathcal{C}^{\text{i.o.}}|R(\Delta)) = 1$. □

**Corollary 4.14.** *Let $\mathcal{C}$ be a language class that is closed under intersection and contains the trivial language $A = \varnothing$. Then Hausdorff dimension $\dim_H(\mathcal{C}^{\text{i.o.}}) \geq 1/2$ and packing dimension $\dim_P(\mathcal{C}^{\text{i.o.}}) = 1$.*

*Proof.* Let $\Delta$ be all functions from $\{0,1\}^* \to \{0,1\}^*$. This follows immediately from Theorem 4.13. □

Now by Observation 3.1, we have the following corollary.

**Corollary 4.15.** *For all $c > 0$*

$$\dim_{\text{p}_2}(\text{SIZE}^{\text{i.o.}}(n^c)) = \dim(\text{SIZE}^{\text{i.o.}}(n^c)|\text{EXP}) = 1/2,$$

$$\dim_{\text{p}_3}(\text{P/poly}^{\text{i.o.}}) = \dim(\text{P/poly}^{\text{i.o.}}|\text{E}_3) = 1/2$$

*and*

$$\text{Dim}_{\text{p}_3}(\text{P/poly}^{\text{i.o.}}) = \text{Dim}(\text{P/poly}^{\text{i.o.}}|\text{E}_3) = 1.$$

By Theorem 3.2, 0th scale is the best scale for evaluting scaled $p_3$-dimension of $P/poly^{i.o.}$. We cannot obtain more informative strong dimension results about $P/poly^{i.o.}$ and it is not hard to show that for any infinitely often class, the scaled strong dimension is 1 for every scale $g_i$ (even for $i < 0$, see [7]).

**Acknowledgment.** I thank Jack Lutz for extremely helpful discussions and tremendous amount of time he spent helping me doing research. I thank John Hitchcock for very helpful comments and early access to [8]. I also thank Satyadev Nandakumar, Anumodh Abey and Fengming Wang for their help in improving the presentation.